%% file: main.tex
\PassOptionsToPackage{table}{xcolor}
\documentclass[acmsmall,screen,nonacm]{acmart}

\usepackage{etoolbox}
\usepackage{textgreek}
\usepackage{tcolorbox}

\AtBeginDocument{%
  }

\copyrightyear{2026}
\acmYear{2026}

\newcommand{\codeurl}{https://github.com/aliabigdeli/llm-falsifier}

\begin{document}

\title{Large Language Models as Falsifiers for Cyber-Physical Systems}

\author{Ali ArjomandBigdeli}
\email{aarjomandbig@cs.stonybrook.edu}
\orcid{0009-0003-7758-6839}
\affiliation{%
  \institution{Stony Brook University}
  \city{Stony Brook}
  \state{New York}
  \country{USA}
}

\author{Jiawei Zhou}
\email{jiawei.zhou.1@stonybrook.edu}
\orcid{0000-0001-5590-6270}
\affiliation{%
  \institution{Stony Brook University}
  \city{Stony Brook}
  \state{New York}
  \country{USA}
}

\author{Stanley Bak}
\email{stanley.bak@stonybrook.edu}
\orcid{0000-0003-4947-9553}
\affiliation{%
  \institution{Stony Brook University}
  \city{Stony Brook}
  \state{New York}
  \country{USA}
}

%% Short author list for page headers.
\renewcommand{\shortauthors}{ArjomandBigdeli et al.}

\begin{abstract}
\input{sections/abstract}
\end{abstract}




\maketitle

\input{sections/intro}

\input{sections/background}

\input{sections/related}

\input{sections/methodology}

\input{sections/evaluation}

\input{sections/limitations}

\input{sections/conclusion}

%% Acknowledgments.
\begin{acks}
This material is based upon work supported by the National Science Foundation
under Award No.\ 2237229 and 2448869.
\end{acks}

%% Bibliography
\bibliographystyle{ACM-Reference-Format}
\bibliography{ref}

\end{document}

%% file: sections/abstract.tex
Falsification searches for counterexamples to formal specifications in cyber-physical systems (CPS). With specifications written in Signal Temporal Logic (STL), falsification can be formulated as a robustness optimization problem, traditionally tackled with black-box search algorithms. In parallel, large language models (LLMs) have recently emerged as surprisingly effective optimizers when coupled with iterative prompting. In this work, we connect these ideas and introduce \textsc{LLM-Falsifier}, an LLM-based approach that falsifies specifications by minimizing the STL robustness degree. Beyond generic prompt-based optimization, our key idea is to expose the LLM to semantic information that is natural for language models but absent from standard numerical optimizers, including natural-language input and output names, output trajectories, and critical-time witnesses for the minimum robustness value. These additions enable smarter and more sample-efficient robustness search. On the ARCH-COMP falsification benchmarks, \textsc{LLM-Falsifier} is shown to outperform existing falsification tools based on a range of optimization paradigms, from surrogate-based and Bayesian optimization to search-based testing, on 14 of 21 specifications when measured by the average number of simulations required to find a counterexample.

%% file: sections/intro.tex
\section{Introduction}
\label{sec:intro}

Falsification is a search for errors in cyber-physical system (CPS) designs. Given a model (typically a Simulink or Python simulator) and a specification expressed in Signal Temporal Logic (STL)~\cite{maler2004monitoring}, a falsification algorithm tries to find an initial state and input signal that cause the model to violate the specification.
Such specifications detail expected requirements for the dynamical system, such as achieving time-sensitive goals, preventing unsafe behaviors, and ensuring desirable behaviors over specific time periods.

Classical logics have Boolean semantics, where a formula is either true or false. STL can also be interpreted with quantitative semantics that evaluate how strongly a signal satisfies or violates a specification~\cite{donze2010robust}.
This quantitative interpretation, known as the \emph{robustness degree} or \emph{robustness value}, assigns a real number that quantifies the margin of satisfaction or violation~\cite{fainekos2009robustness,rizk2009general}.
This allows a given specification and a system trajectory (i.e., a ``signal'') to be systematically transformed into a single scalar robustness value, where the sign denotes whether the specification is satisfied or violated, and the magnitude captures how strongly that satisfaction or violation occurs. This key property enables STL to frame the falsification problem as a numeric optimization problem where the goal is to minimize the robustness value.

\begin{figure*}[t]
  \centering
  \includegraphics[width=0.9\textwidth]{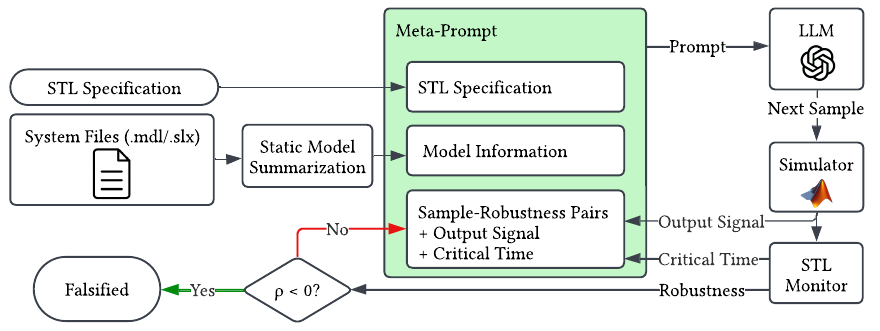}
  \caption{\textsc{LLM-Falsifier} prompts a large language model to generate samples for a CPS falsification problem.}
  \Description{Block diagram of the closed-loop LLM-Falsifier workflow. The STL specification and system files feed a static model summarization step that builds a meta-prompt. The LLM proposes a candidate input, the simulator runs it, and an STL monitor computes the robustness value and critical time. If the robustness is negative the loop terminates with a counterexample; otherwise the sample and its feedback are appended to the prompt history and the loop repeats.}
  \label{fig:teaser}
\end{figure*}

The robustness measure in STL is non-smooth and non-convex due to the nested min/max operators in its definition.
Traditional falsification approaches solve this problem using black-box, derivative-free optimization strategies.
Recently, large language models (LLMs) have also been shown to be capable of solving derivative-free optimization problems using an iterative prompting technique called Optimization by PROmpting (OPRO)~\cite{yang2024large}.
In OPRO, a meta-prompt combines a natural-language description of the problem with previously evaluated solution--score pairs, and at each optimization step the LLM is prompted to generate several new candidate solutions.
These candidates are scored outside the LLM, and only the best-scoring solutions found so far are kept in the meta-prompt, sorted by score, for the next step.
In contrast to traditional optimization methods, OPRO therefore uses natural-language prompts to iteratively propose candidate solutions from the problem description.

In this work, we connect these two ideas and introduce \textsc{LLM-Falsifier}, which is, to our knowledge, the first LLM-driven robustness-guided falsifier for CPS.
Our central claim is that LLMs can directly perform falsification by optimizing robustness values, and that they become substantially more effective when the search process is expressed using natural-language-based semantic information that they are naturally good at exploiting.
We draw inspiration from OPRO, but both our problem and our search loop differ. Instead of focusing solely on numerical search and relying on a single scalar score for each sample, we enrich the process with semantic and contextual feedback that the LLM can reason about. This feedback includes natural-language names for inputs and outputs, output signal values over time, and a \emph{critical time point} that witnesses the minimum robustness value.
The critical time point is the time at which the final robustness value exhibits sensitivity to changes in the signal, and we show that including the output signal values at this time in the prompt improves LLM-driven falsification performance.
Together, these components enable the model to leverage semantic comprehension and causal reasoning more effectively, introduce a strong inductive bias, and guide exploration using informative heuristics that accelerate falsification.
Beyond the content of the prompt, the search loop itself also differs from OPRO: because every candidate input must be evaluated by a simulation, our loop starts without any initial samples, requests a single sample per iteration, and involves no external ranking or selection among candidates (Section~\ref{sec:overview}).

The proposed closed-loop falsification workflow is shown in Figure~\ref{fig:teaser}: given the \textbf{STL Specification} and the \textbf{System Files}, a \textbf{Static Model Summarization} step extracts model information such as input and output signal names to fill in a \textbf{Meta-Prompt}, the \textbf{LLM} proposes the next sample, a \textbf{Simulator} runs it, and an \textbf{STL Monitor} computes the robustness value $\rho$ together with its critical time.
If $\rho<0$, a counterexample has been found; otherwise the sample, its robustness, and the associated output-signal and critical-time information are appended to the prompt as \textbf{Sample-Robustness Pairs}, and the process repeats (Section~\ref{sec:overview}).

We evaluate \textsc{LLM-Falsifier} on the Applied Verification for Continuous and Hybrid Systems competition (ARCH-COMP) benchmarks. The results demonstrate state-of-the-art performance, with our approach often outperforming specialized falsification tools based on classical optimization algorithms.
Throughout the paper, \emph{sample efficiency} refers to the number of candidate inputs that must be generated and simulated before a counterexample is found. This is the primary cost measure in falsification, where every sample requires an expensive simulation, and it is the quantity reported by the ARCH-COMP evaluation protocol.
On six specifications, the LLM typically finds a falsifying input on the very \emph{first} simulation, an outcome that is essentially impossible for a purely numerical optimizer, which has no information before its first sample.
We further study the effect of the underlying model and reasoning effort, including an open-source model, and inspect the reasoning traces of the LLM to understand how it constructs falsifying inputs.

The main contributions of this paper are as follows:
\begin{itemize}
    \item We introduce \textsc{LLM-Falsifier}, a method that lets an LLM directly perform CPS falsification by iteratively proposing inputs that optimize STL robustness values.
    \item We show that LLMs can exploit semantic information, such as natural-language variable names, output trajectories, and critical-time witnesses, to conduct smarter and more sample-efficient robustness optimization.
    \item We evaluate \textsc{LLM-Falsifier} on the ARCH-COMP 2025 benchmarks and show that it outperforms widely adopted falsification tools based on a range of optimization paradigms, from surrogate-based and Bayesian optimization to search-based testing, on 14 out of 21 specifications.
\end{itemize}
The implementation of \textsc{LLM-Falsifier} and the scripts used in our experiments are publicly available.\footnote{\url{\codeurl}}

This paper is organized as follows.
Section~\ref{sec:background} provides background on the STL falsification problem, Section~\ref{sec:related} discusses related work, and Section~\ref{sec:method} introduces our methodology, describing the LLM-based falsification workflow and its enhancements with natural-language signal names, output signal values, and critical time points.
Section~\ref{sec:eval} presents our experimental evaluation on the ARCH-COMP falsification benchmarks, including an ablation study quantifying the effect of each enhancement and a comparison of LLMs.
Section~\ref{sec:limitations} discusses limitations and Section~\ref{sec:conclusion} concludes with a summary of findings.
Throughout, we use the ARCH-COMP Automatic Transmission benchmark and its specification AT1 as a running example, introduced in Section~\ref{sec:background} and followed through the method and the evaluation.

%% file: sections/background.tex
\section{Background}
\label{sec:background}

Let $\mathbb{T} \subseteq \mathbb{R}_{\ge 0}$ denote the time domain (typically $[0,T_{\mathrm{end}}]$). In CPS falsification, a simulator (model) is a function $\mathcal{M}$ that, given an initial state $x_0$ and an input signal $u: \mathbb{T} \to \mathbb{R}^{n_i}$, produces an output signal $\mathbf{x}: \mathbb{T} \to \mathbb{R}^{n_o}$, with $\mathbf{x}(\cdot) = \mathcal{M}(x_0,u(\cdot))$.

The syntax of (bounded) STL is given by the grammar
\[
\varphi ::= \text{true} \;\mid\; \mu  \;\mid\; \neg\varphi \;\mid\; \varphi_1\wedge\varphi_2 \;\mid\; \varphi_1 \;\mathbf{U}_{[a,b]}\; \varphi_2,
\]
where the atom $\mu \equiv h(\mathbf{x}(t)) \geq 0$ and $0\le a\le b$ are real time bounds. From the \emph{until} operator $\mathbf{U}_{[a,b]}$ we can derive the bounded temporal operators $\mathbf{F}$ (eventually) and $\mathbf{G}$ (globally):
\[
\mathbf{F}_{[a,b]}\varphi \;=\; \mathrm{true}\;\mathbf{U}_{[a,b]}\;\varphi,\ \mathbf{G}_{[a,b]}\varphi \;=\; \neg\mathbf{F}_{[a,b]}\neg\varphi.
\]

STL admits a quantitative semantics that assigns to every triple $(\varphi,\mathbf{x},t)$ a \emph{robustness degree} in $\mathbb{R} \cup \{\infty, -\infty\}$ denoted $\rho(\varphi,\mathbf{x},t)$. The robustness encodes both satisfaction and the margin of satisfaction: by convention $\rho(\varphi,\mathbf{x},t)>0$ indicates satisfaction at time $t$, $\rho(\varphi,\mathbf{x},t)<0$ indicates violation, 
and the magnitude $|\rho(\varphi,\mathbf{x},t)|$ measures how strongly $\varphi$ is satisfied or violated. We use the standard min/max-based definition of robustness semantics~\cite{maler2004monitoring,fainekos2012verification}:
\begin{align}
\rho(\text{true},\mathbf{x},t) &\;=\; \infty, \label{eq:rob-true}\\
\rho(\mu,\mathbf{x},t) &\;=\; h(\mathbf{x}(t)), \label{eq:rob-pred}\\
\rho(\neg\varphi,\mathbf{x},t) &\;=\; -\,\rho(\varphi,\mathbf{x},t), \\
\rho(\varphi_1\wedge\varphi_2,\mathbf{x},t) &\;=\; \min\big(\rho(\varphi_1,\mathbf{x},t),\ \rho(\varphi_2,\mathbf{x},t)\big).
\end{align}

For a time-bounded \emph{until} formula, $\psi \equiv \varphi_1\ \mathbf{U}_{[a,b]}\ \varphi_2$, the standard quantitative semantics is
\begin{equation}\label{eq:rob-until}
\rho(\psi,\mathbf{x},t)
= \sup_{t'\in t+[a,b]} \min\Big(\rho(\varphi_2,\mathbf{x},t'),\
\inf_{t''\in[t,t']}\rho(\varphi_1,\mathbf{x},t'')\Big).
\end{equation}
From this definition, one can derive simplified robustness formulas for the \textbf{F} (eventually) and \textbf{G} (globally) operators:
\begin{align}
\rho(\mathbf{F}_{[a,b]}\varphi,\mathbf{x},t) &= \sup_{t'\in t+[a,b]}\rho(\varphi,\mathbf{x},t'),\\
\rho(\mathbf{G}_{[a,b]}\varphi,\mathbf{x},t) &= \inf_{t'\in t+[a,b]}\rho(\varphi,\mathbf{x},t'). \label{eq:rob-G}
\end{align}

\noindent
\textbf{Problem Statement (STL Falsification):} Given a model $\mathcal{M}$ and STL formula $\varphi$, the falsification problem is to find an initial condition $x_0$ and admissible input signal $u(\cdot)$ such that the resulting trajectory $\mathbf{x}=\mathcal{M}(x_0,u)$ violates $\varphi$. 

Using quantitative semantics, falsification is commonly cast as the following optimization problem:
\begin{equation}\label{eq:falsification-problem}
\min_{(x_0,u)}\ \rho(\varphi,\mathcal{M}(x_0,u),0).
\end{equation}
A solution with objective value $\rho(\varphi,\mathcal{M}(x_0,u),0)<0$ constitutes a counterexample (witness) that falsifies the specification.

The nested use of $\min$, $\sup$, and $\inf$, as well as the complexity of the CPS simulation model $\mathcal{M}$, make the robustness function in general non-smooth and non-convex, which in turn motivates the widespread use of derivative-free and heuristic optimization methods in falsification~\cite{fainekos2012verification,khandait2025arch}.
The search problem is often simplified by constraining the input signal using a finite parameterization, for example, the input may be restricted to be a piecewise-linear interpolation between values given at evenly spaced time points.
This is the standard approach in tools such as Breach~\cite{donze2010breach} and S-TaLiRo~\cite{annpureddy2011s}. 

\begin{figure}[!t]
  \centering
  \includegraphics[width=0.85\linewidth]{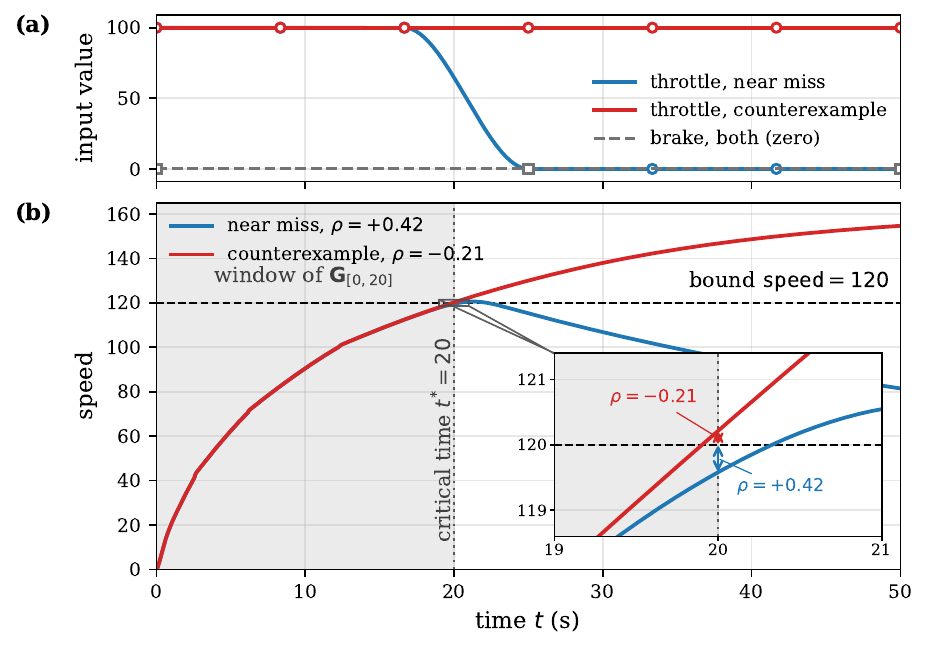}
  \caption{The running example.
  (a)~Two candidate inputs for the Automatic Transmission model, each given by $7$ throttle and $3$ brake control points (markers) that are interpolated in between: full throttle at the first three control points and none afterwards (blue), and full throttle throughout (red); the brake is zero in both.
  (b)~The resulting speed traces, the bound $\texttt{speed}\le 120$ of $\varphi_{\mathrm{AT1}}$ (dashed), and its time window $[0,20]$ (shaded).
  In both cases the supremum of the speed over the window is attained at $t^{*}=20$ (dotted), so the robustness is the signed distance from the trace to the bound at $t^{*}$ (inset): $+0.42$ for the blue trace, which stays below the bound inside the window and exceeds it only afterwards, and $-0.21$ for the red trace, which is a counterexample.}
  \Description{Two stacked plots over a fifty second horizon. The top plot shows the throttle input of two samples: one drops smoothly from one hundred to zero between seventeen and twenty-five seconds, the other stays at one hundred; the brake is zero for both. The bottom plot shows the resulting vehicle speed traces with a dashed horizontal bound at one hundred twenty and a shaded window from zero to twenty seconds. Both traces reach about one hundred twenty at twenty seconds. An inset zooms in around that time and shows that one trace is slightly below the bound, robustness plus zero point four two, and the other slightly above it, robustness minus zero point two one.}
  \label{fig:at1-example}
\end{figure}

\paragraph{Running example}
To make these definitions concrete, we use the \emph{Automatic Transmission} (AT) benchmark~\cite{hoxha2014benchmarks} from the ARCH-COMP falsification competition~\cite{khandait2025arch} as a running example throughout the paper.
The model $\mathcal{M}$ is a Simulink model of a vehicle with an automatic transmission, simulated over the horizon $\mathbb{T}=[0,50]$ seconds.
It has two input signals, the \texttt{throttle} position with range $[0,100]$ and the \texttt{brake} with range $[0,325]$, and three output signals: the vehicle \texttt{speed}, the engine speed \texttt{rpm}, and the selected \texttt{gear}.
The initial state is fixed (the vehicle starts at rest), so the search is over the input signal $u(t)=(\texttt{throttle}(t),\texttt{brake}(t))$ only.
As the specification of the running example we use the AT1 specification of the benchmark,
\[
\varphi_{\mathrm{AT1}} \;=\; \mathbf{G}_{[0,20]}\,(\texttt{speed} \le 120),
\]
which states that the vehicle speed must not exceed $120$ during the first $20$ seconds.
Its only atom is $\mu \equiv h(\mathbf{x}(t))\ge 0$ with $h(\mathbf{x}(t)) = 120-\texttt{speed}(t)$, so by Eqs.~\eqref{eq:rob-pred} and~\eqref{eq:rob-G} the robustness of a trajectory is
\[
\rho(\varphi_{\mathrm{AT1}},\mathbf{x},0) \;=\; \inf_{t\in[0,20]}\big(120-\texttt{speed}(t)\big) \;=\; 120-\sup_{t\in[0,20]}\texttt{speed}(t),
\]
i.e., the margin by which the peak speed in the first $20$ seconds stays below $120$.
It is positive as long as the speed stays below the bound and becomes negative exactly when the speed exceeds $120$ somewhere in the window.
Figure~\ref{fig:at1-example}(b) shows two such trajectories.
The first peaks at $119.6$ within the window, so its robustness is $\approx 0.4$ and it does not violate the specification, even though its speed exceeds $120$ shortly after $t=20$, outside the window; the second reaches $120.2$ at $t=20$, has robustness $\approx -0.2$, and constitutes a counterexample.
Following the signal parameterization that \textPsi-TaLiRo uses for ARCH-COMP Instance~1 (Section~\ref{ssec:eval_setup}), the throttle signal is given by $7$ control points and the brake signal by $3$ control points, evenly spaced over the $50$-second horizon and interpolated in between.
A candidate input is therefore a vector in $[0,100]^7\times[0,325]^3$, and the falsification problem in Eq.~\eqref{eq:falsification-problem} for $\varphi_{\mathrm{AT1}}$ becomes a $10$-dimensional box-constrained search for throttle and brake control points whose interpolated signals drive the speed above $120$ before $t=20$ seconds.
The two inputs of Figure~\ref{fig:at1-example}(a) are such vectors: full throttle at the first three control points and none afterwards, $(100,100,100,0,0,0,0,\,0,0,0)$, produces the near miss, whereas full throttle at all seven control points, $(100,\dots,100,\,0,0,0)$, produces the counterexample; the brake is zero in both.
Section~\ref{sec:method} shows how \textsc{LLM-Falsifier} performs this search, and Section~\ref{sec:eval} reports how many simulations it needs (the AT1 rows of Tables~\ref{tab:archcomp-rank}--\ref{tab:model-comparison}).

%% file: sections/related.tex
\section{Related Work}
\label{sec:related}

Correctness is especially important in CPS, as mistakes can have real-world consequences. While formal approaches like model checking~\cite{clarke1994model} and deductive verification~\cite{owre1996pvs} continue to make progress in CPS domains~\cite{fulton2015keymaera}, the complexity of CPS, the ad-hoc nature of engineering in practice, and the fundamental undecidability of the verification problem for many practical cases~\cite{henzinger1995s} limit the applicability of these methods. Test-driven methods like falsification have been developed as practical alternatives with wider applicability at the cost of less rigorous guarantees~\cite{kapinski2016simulation}.
Classical CPS falsification is largely framed as black-box, derivative-free optimization over the STL robustness landscape, with mature toolchains such as Breach~\cite{donze2010breach} and S-TaLiRo~\cite{annpureddy2011s} providing simulation, monitoring, and a search backbone. The ARCH-COMP falsification category~\cite{khandait2025arch} consolidates a diverse set of strategies on top of this foundation, including surrogate-based methods (ARIsTEO, FlexiFal, FReaK), Bayesian optimization (\textPsi-TaLiRo's ConBO-LS), search-based testing (ATheNA), automata learning (FalCAuN), and Monte Carlo Tree Search (ForeSee); per-tool descriptions are given in Section~\ref{ssec:eval_setup}. Our work differs from all of these by using a large language model itself as the search engine, exploiting semantic signal names, output trajectories, and critical-time witnesses that classical numerical optimizers cannot consume.

STL is also used beyond falsification, for example in planning~\cite{mehdipour2019arithmetic}, robotics~\cite{haghighi2019control}, automotive~\cite{fainekos2012verification} and multi-agent systems~\cite{pant2018fly}, as well as for monitoring~\cite{donze2013efficient,deshmukh2017robust,bartocci2018specification} and specification mining~\cite{jin2013mining}.
Within numerical falsification, the formulation of the robustness measure determines the complexity of the optimization. Options beyond the min/max semantics used in this work include arithmetic-geometric mean robustness~\cite{mehdipour2019arithmetic}, MIP formulations~\cite{belta2019formal}, and smooth cumulative semantics~\cite{haghighi2019control}, which replace hard min/max operations with smooth aggregates suited to gradient-based optimization and control synthesis. They may also benefit LLM-driven falsification, but the notion of critical time would need to be reconsidered if the robustness semantics are adjusted.
Our critical time points, which identify times in the simulation \emph{output} signals, are related to trace diagnostics for STL, which have been used for fault localization in Simulink/Stateflow models~\cite{bartocci2018localizing} and to extend robustness to distinguish between input and output signals~\cite{ferrere2019interface}, building on earlier diagnostics for LTL and MTL~\cite{ferrere2015trace}. Other work studies the harder problem of identifying the times and values of \emph{input} signals responsible for a counterexample~\cite{diwakaran2017analyzing}. Rather than providing complete trace diagnostics in the prompt, we select a single witness time and its associated signal values as a compact guide for the search process.

Large language models have recently been used as general-purpose decision-making and search modules through in-context learning (ICL), where the model is conditioned on natural-language instructions and a small set of examples rather than updated through gradient-based training. This capability became especially prominent with GPT-3~\cite{brown2020language} and has since been explored in several sequential decision-making settings, including reinforcement learning~\cite{laskin2022context,monea2024llms}. More broadly, these efforts reflect the growing use of foundation models to support engineering workflows~\cite{yuksel2023review}. Closely related to our setting, Optimization by PROmpting (OPRO)~\cite{yang2024large} uses an LLM as an iterative, derivative-free optimizer by prompting it with an optimization problem description and previously evaluated solution--score pairs.
At each optimization step, the meta-prompt is used to generate several new candidate solutions, which are then evaluated and ranked outside the LLM, with only the best-scoring solutions retained (and sorted by score) in the next meta-prompt. This strategy can be viewed as a form of hill climbing: the LLM proposes multiple candidates (exploration), and a small subset is selected outside the LLM for further improvement (exploitation).
Our work addresses a related but different challenge, namely the combination of numerical and linguistic reasoning for optimization in falsification, where each candidate evaluation is a simulation; \textsc{LLM-Falsifier} therefore needs no initial samples, generates a single sample per iteration, and does not rely on an external selection strategy (Section~\ref{sec:overview}).

Within software engineering, LLMs have also been investigated for testing-related tasks such as test generation and fuzzing~\cite{xia2024fuzz4all,oliinyk2024fuzzing}% [ICCPS] oliinyk
, and for constructing counterexamples that refute incorrect programs~\cite{sinha2025can}.
These works suggest that LLMs can help propose informative inputs by exploiting structure expressed in natural language, code, or prior execution feedback.
In the CPS domain, LLMs have been proposed as a component of formal testing pipelines for learning-enabled systems~\cite{zheng2024testing} and are widely used to generate test scenarios for automated driving~\cite{zhao2026survey}, typically operating on high-level scenario descriptions with Boolean pass/fail outcomes rather than continuous input signals scored by a robustness monitor.
However, CPS falsification differs fundamentally from conventional software fuzzing. In many software-testing settings, executions are relatively cheap and the main challenge is to generate diverse, high-value seeds. In contrast, falsification typically relies on expensive simulations of dynamical systems, and each query is evaluated through a robustness objective derived from an STL specification. As a result, the search must be much more sample-efficient and tightly coupled to quantitative feedback from the monitor.

To the best of our knowledge, this is the first study to employ an LLM directly as a robustness-guided falsifier for CPS. We demonstrate that LLMs already have the reasoning ability required for counterexample search in falsification, and that this ability can be strengthened by integrating semantic and numerical feedback.

%% file: sections/methodology.tex
\section{Methodology}
\label{sec:method}
This section presents our LLM-driven approach for CPS falsification. Section~\ref{sec:overview} describes the closed-loop \textsc{LLM-Falsifier} architecture. Section~\ref{sec:signal_names} then develops a sequence of four progressively enriched meta-prompt variants (MP1--MP4) that expose increasing amounts of semantic context to the LLM, and Section~\ref{sec:critical-time} formalizes the critical-time witness used by the most enriched variant.

\subsection{Overview}
\label{sec:overview}
The starting point is the STL falsification problem in Eq.~\eqref{eq:falsification-problem}. \textsc{LLM-Falsifier} addresses it by treating a large language model as the derivative-free optimizer: at each iteration, the LLM proposes a new candidate input from a prompt that summarizes the history of past samples and their robustness values. Figure~\ref{fig:teaser} shows the closed-loop workflow. Given the \textbf{STL Specification} and the \textbf{System Files} (e.g., a Simulink model), a \textbf{Static Model Summarization} step extracts static model information for the \textbf{Meta-Prompt}. The \textbf{LLM} proposes the next sample, the \textbf{Simulator} runs it, and an \textbf{STL Monitor} returns the robustness value $\rho$ along with its witness time, which we call the \emph{critical time}. If $\rho<0$, the process terminates; otherwise the sample and its robustness, together with any additional per-sample feedback we choose to include (defined in Section~\ref{sec:signal_names}), are appended to the prompt as \textbf{Sample-Robustness Pairs}, and the loop repeats.
The history is empty in the first iteration, so the first simulated input is already proposed by the LLM; each iteration requests exactly one sample and therefore costs exactly one simulation, and the prompt retains the ten most recent samples in chronological order without any score-based selection or reordering.
What concretely populates the Meta-Prompt and the Sample-Robustness Pairs defines the prompt variant; we develop these variants next.

\paragraph{Running example}
For the AT running example of Section~\ref{sec:background}, one iteration of the loop proceeds as follows.
The Static Model Summarization step parses the Simulink model files and recovers the input names \texttt{throttle} and \texttt{brake} with their ranges, and it extracts the output name \texttt{speed} from the specification $\varphi_{\mathrm{AT1}}$, since that is the signal the specification constrains; together with the specification itself and the signal parameterization ($7$ throttle and $3$ brake control points), this information populates the Meta-Prompt.
Figure~\ref{fig:meta-prompt} shows the resulting prompt.
The LLM replies with a single candidate point, i.e., ten numbers giving the throttle and brake values at the control points, which we parse from its response.
The Simulator interpolates these values into continuous input signals, runs the Simulink model for $50$ seconds, and returns the output trace, from which the STL Monitor evaluates $\rho(\varphi_{\mathrm{AT1}},\mathbf{x},0)=120-\sup_{t\in[0,20]}\texttt{speed}(t)$ together with its critical time, the time in $[0,20]$ at which the speed is largest.
For the first history entry shown in Figure~\ref{fig:meta-prompt}, which is the near miss of Figure~\ref{fig:at1-example}, the speed peaked at $119.6$ (rounded in the prompt) at $t=20$ seconds, so $\rho=0.419>0$ and the specification is not yet violated: the point, its robustness, the sampled speed trajectory, and the critical-time witness are appended to the history, and the LLM is prompted again.
The search stops as soon as a proposed point yields $\rho<0$, i.e., a throttle and brake profile under which the speed exceeds $120$ within the first $20$ seconds.
Section~\ref{ssec:explainability} shows the reasoning with which the LLM arrives at such a point for this example, and the AT1 rows of Tables~\ref{tab:archcomp-rank}--\ref{tab:model-comparison} report how many simulations this takes.

\subsection{Semantic Information Enhancement}
\label{sec:signal_names}

\begin{figure}[!t]
\centering
\begin{tcolorbox}[
    colback=green!10!white, % Background color
    colframe=green!50!gray, % Frame color
    coltitle=black!80,      % Title text color
    fonttitle=\bfseries,    % Title font
    fontupper=\small\ttfamily, % Box content font
    title=Example of our Prompt for the AT System,
    rounded corners,
    boxrule=0.8pt,
    left=5pt,right=5pt,top=5pt,bottom=5pt
]
You are an optimization assistant for a system falsification task. Your goal is to find input parameters that violate system specifications (negative robustness values indicate violations). \\[6pt]
\textcolor{violet}{SYSTEM SPECIFICATION TO VIOLATE:}

\textcolor{violet}{STL Formula: G[0, 20] (speed <= 120)} \\[6pt]
\textcolor{violet}{OUTPUT VARIABLES:}\\
\textcolor{violet}{speed} \\[6pt]
INPUT CONTROL PARAMETERS:
The input space has 10 dimensions representing control parameters:

Dimension 1: [0.0, 100.0] - \textcolor{violet}{Throttle level at t=0.0s}

\vdots
Dimension 7: [0.0, 100.0] - \textcolor{violet}{Throttle level at t=50.0s}

Dimension 8: [0.0, 325.0] - \textcolor{violet}{Brake pressure at t=0.0s}

\vdots
Dimension 10: [0.0, 325.0] - \textcolor{violet}{Brake pressure at t=50.0s} \\[6pt]

OPTIMIZATION CONTEXT:\\
- Lower robustness values are better (negative values indicate specification violations)\\
- Use the semantic meaning of each dimension and the specification requirements to make informed decisions \\
- \textcolor{blue}{Pay attention to the system output states to understand how inputs affect system behavior}\\[6pt]
RECENT OPTIMIZATION HISTORY (last 10 samples):

Sample 1: ['100.000', \dots , '0.000'] -> Robustness: 0.419\\
    \textcolor{blue}{Output of Sample 1: speed: at 0.0s = 0.0, at 8.3s = 82.5,\dots , at 50.0s = 81.5;} \textcolor{cyan}{Robustness value of 0.419 for this sample originates from state values at time = 20.0s: speed = 119.6}

\vdots

Based on this history and the specification requirements, generate a new sample point that is different from all points above and is likely to achieve a lower robustness value 
(violation).
Consider:\\
1. Which parameter combinations led to lower robustness values in the history\\
2. The physical/logical meaning of each parameter and how it affects the output variables\\
\textcolor{violet}{3. How the specification constrains the output variables and what inputs might violate these constraints}\\
\textcolor{blue}{4. How the system outputs changed with different input parameters}\\
\textcolor{blue}{5. Patterns in the output states that might indicate approaching or achieving violations}
%Think step by step \dots end with '\textless/point\textgreater'.
\end{tcolorbox}
\caption{Enriched meta-prompt (MP4) for the specification AT1 of the Automatic Transmission benchmark; removing the colored augmentations yields the baseline prompt MP1.}
\Description{Text box showing the full meta-prompt sent to the LLM for the Automatic Transmission benchmark. It contains the STL specification to violate, named output variables, named input control dimensions with ranges, optimization guidance, and a history of past samples annotated with robustness values, output trajectories, and critical-time information. Colored text marks the three progressive prompt augmentations.}
\label{fig:meta-prompt}
\end{figure}

We start from a minimal instantiation of the loop, which we call \textbf{Meta-Prompt~1 (MP1)}. In MP1, input dimensions are listed by index with no natural-language names, no specification, and no output information; the per-sample feedback in the prompt history is the scalar robustness only. MP1 corresponds to Figure~\ref{fig:meta-prompt} with all colored augmentations removed, and represents our simplest (least enriched) meta-prompt for the falsification problem in Eq.~\eqref{eq:falsification-problem}, i.e., numeric solution--score pairs only, within the loop of Section~\ref{sec:overview}.

This baseline treats falsification as a purely numeric, black-box search where input dimensions are listed by index with no semantic information. 
This forgoes a potential advantage of LLM-based optimization, the ability to reason about the semantics and causality within a model. 
To exploit those capabilities, we modify the meta-prompt to include more context. 
Concretely, we include (i) the natural language names of the inputs and outputs, (ii) the output state signal values, and (iii) a single witness time that we call the \emph{critical time point}, whose signal values are relevant to the computed robustness score. 
These modifications serve to turn a pure numerical search into a contextual reasoning task that the LLM can potentially solve more effectively.

An example of a prompt, slightly modified for brevity, containing these enhancements is shown in Figure~\ref{fig:meta-prompt}.
The prompt corresponds to the one used to falsify the specification AT1 of the 
\textbf{Automatic Transmission} (AT) benchmark~\cite{hoxha2014benchmarks} from the ARCH-COMP~\cite{khandait2025arch} competition.
The colors correspond to the different levels of proposed enhancements.

% 
% \noindent 
\textbf{Specification context (shown in \textcolor{violet}{violet})} includes the STL formula of the requirement to be violated. Providing this explicit objective (for example, \texttt{G[0,20](speed <= 120)}) directs the LLM toward input changes that influence the specification-relevant output signals. Additionally, we automatically extract the relevant output signal name from the STL specification and explicitly include it in the prompt. Beyond including the specification itself, we provide explicit guidance instructing the model to leverage it to encourage logical reasoning.
Since the STL specification references signal names, we also give each input dimension a text name associated with its physical meaning, such as throttle or brake values at specific times, along with the relevant output variables (e.g., \texttt{speed}). This enables the LLM to reason about causal relationships, such as inferring that increasing the throttle input may increase vehicle speed output.
This augmentation defines \textbf{Meta-Prompt~2 (MP2)}, which extends the baseline prompt (MP1) with the \textcolor{violet}{violet} specification and signal-name context shown in Figure~\ref{fig:meta-prompt}.

% 
% \noindent
\textbf{Output-signal feedback (shown in \textcolor{blue}{blue})} expands the optimization history beyond scalar robustness scores by including the output state signal trajectories.
Since full output traces are large and could bloat the prompt, we include a compact representation instead: pointwise output state values sampled at the same times as the input control points. This keeps the prompt concise while preserving the key information the LLM needs to reason about cause and effect. When the prompt includes output signal values, we further include explicit instructions encouraging the model to leverage this information, as shown in the figure. Adding this \textcolor{blue}{blue} output-signal context on top of MP2 defines \textbf{Meta-Prompt~3 (MP3)}.

% 
% \noindent
\textbf{Critical time points (shown in \textcolor{cyan}{cyan})} add to each history sample a computed time point and the signal values at that time, which serve as a witness to the robustness value.
The exact definition and details of this calculation will be explained next in Section~\ref{sec:critical-time}. Adding this \textcolor{cyan}{cyan} critical-time information on top of MP3 yields \textbf{Meta-Prompt~4 (MP4)}, our full prompt variant.

These refinements preserve the baseline \textsc{LLM-Falsifier} loop while potentially leveraging model domain information (signal names, output state values and critical point) to bias the LLM toward semantically meaningful searches. In summary, MP1 is the baseline prompt, MP2 adds specification and signal-name context, MP3 additionally includes output signal values, and MP4 further adds critical-time information. In our ablation study, we show that this leads to more focused samples and faster discovery of counterexamples.

\subsection{Critical Time Points}
\label{sec:critical-time}
In order to better focus the falsification search, one enhancement was to explicitly identify the times that serve as a witness to the minimum robustness.
%
%
%
%%%%%%%%%%%
To accomplish this, we define the \emph{critical time point} \( \tau(\varphi, \mathbf{x}, t) \in \mathbb{R} \), which identifies the time at which the robustness \( \rho(\varphi, \mathbf{x}, t) \) is realized.
Similar to the quantitative semantics of STL, $\tau$ can be computed recursively based on the syntax of the formula:
\begin{align}
\tau(\text{true}, \mathbf{x}, t) &= t, \label{eq:crit_time_start} \\
\tau(\mu, \mathbf{x}, t) &= t, \\
\tau(\neg \varphi, \mathbf{x}, t) &= \tau(\varphi, \mathbf{x}, t), \\
\tau(\varphi_1 \wedge \varphi_2, \mathbf{x}, t) &=
\begin{cases}
\tau(\varphi_1, \mathbf{x}, t) & \text{if } \rho(\varphi_1, \mathbf{x}, t) \leq \rho(\varphi_2, \mathbf{x}, t) \\
\tau(\varphi_2, \mathbf{x}, t) & \text{otherwise}
\end{cases}
\end{align}
Note that for negation, although the robustness value changes sign, the critical time point remains the same.

The critical time point for the \textbf{F} (eventually) and \textbf{G} (globally) operators can be defined as the time at which the subformula achieves its maximum (for \textbf{F}) or minimum (for \textbf{G}) robustness value.
One complication with these is that if $\varphi$ contains nested temporal operators, it is important to return the critical time point of the \emph{inner} subformula, not the outer one.
\begin{align}
\tau(\mathbf{F}_{[a,b]} \varphi, \mathbf{x}, t) &= \tau(\varphi, \mathbf{x}, t^*) \label{eq:crit_time_F} \\
\text{where } t^* &= \arg\sup_{t' \in t + [a,b]} \rho(\varphi, \mathbf{x}, t')  \nonumber \\[2ex]
\tau(\mathbf{G}_{[a,b]} \varphi, \mathbf{x}, t) &= \tau(\varphi, \mathbf{x}, t^*) \label{eq:crit_time_G} \\
\text{where } t^* &= \arg\inf_{t' \in t + [a,b]} \rho(\varphi, \mathbf{x}, t') \nonumber
\end{align}

\begin{figure}[tbp]
  \centering
  \includegraphics[width=0.73\linewidth]{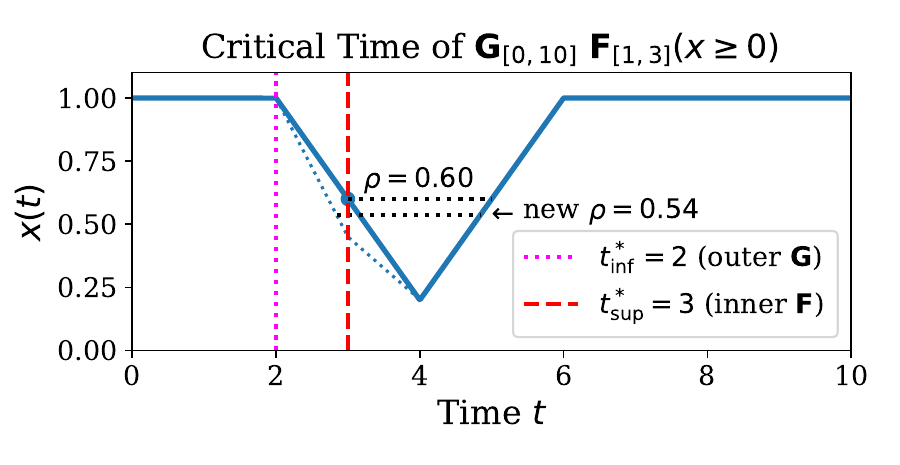}
  \caption{Critical time for $\varphi = \mathbf{G}_{[0,10]}\,\mathbf{F}_{[1,3]}(x \geq 0)$. The critical time $t_{\sup}^*=3$ is determined by the inner temporal operator, not by the outer one ($t_{\inf}^*=2$). Lowering the signal at $t_{\sup}^*$ (dotted blue) reduces the robustness from $0.6$ to $0.54$.}
  \Description{Line plot of a signal over time for the nested formula G over 0 to 10 of F over 1 to 3 of x greater than or equal to 0. The inner critical time at t equals 3 is marked in red and the outer time at t equals 2 in magenta. A dotted blue curve shows that lowering the signal at the inner critical time reduces the overall robustness from 0.6 to 0.54.}
  \label{fig:tau-example}
\end{figure}

When the supremum or infimum in Eqs.~\eqref{eq:crit_time_F} and~\eqref{eq:crit_time_G} is attained at several times, our implementation takes the earliest one.
An illustrative example showing the need to define the critical time point using the inner subformula is shown in Figure~\ref{fig:tau-example} for $\varphi = \mathbf{G}_{[0,10]}\,\mathbf{F}_{[1,3]}(x \geq 0)$.
Here, the critical time corresponding to the solid blue line $x(t)$ is $t^*_{\sup} = 3$ (red vertical line), where the signal $x(t)$ remains low for the next two seconds (the dotted black line is two seconds wide).
The outer time $t^*_{\inf}$ (magenta vertical line) evaluates to time $2.0$.
Put another way, at $t=2$, the robustness of subformula $\mathbf{F}_{[1,3]}(x \geq 0)$ is minimized.
If the signal value was reduced at the critical time, the robustness of the overall formula would decrease.
In the figure, this is shown by the dotted blue line, which represents a modified interpolated output signal, 
slightly reduced at the critical time (the blue dot is moved downward).
The robustness $\rho$ decreases from $\rho=0.6$ with the original solid blue signal to $\rho=0.54$ with the modified dotted blue one.
Of course, it is difficult to precisely manipulate the output signal, but this justifies its use as a reasonable target for the falsifier.

%%%%%%
%
For an until formula, \(\varphi_1 \mathbf{U}_{[a,b]} \varphi_2\), the critical time point corresponds to the inner time associated with either $t^*$ (derived from $\varphi_1$), or $t^{**}$ (derived from $\varphi_2$), depending on which subformula's robustness is smaller:

\begin{align}
&\tau(\varphi_1 \mathbf{U}_{[a,b]} \varphi_2, \mathbf{x}, t) =
\begin{cases}
\tau(\varphi_1, \mathbf{x}, t^*) & \text{if } \rho(\varphi_1, \mathbf{x}, t^*) \leq \rho(\varphi_2, \mathbf{x}, t^{**}) \\
\tau(\varphi_2, \mathbf{x}, t^{**}) & \text{otherwise}
\end{cases} \label{eq:crit_time_end} \\
&t^{*} = \arg\inf_{t'' \in [t, t^{**}]} \rho(\varphi_1, \mathbf{x}, t'') \nonumber \\
&t^{**} = \arg\sup_{t' \in t+[a,b]} \min \left(
\rho(\varphi_2, \mathbf{x}, t'),\;
\inf_{t'' \in [t, t']} \rho(\varphi_1, \mathbf{x}, t'')
\right) \nonumber 
\end{align}
Similar to the earlier temporal operators, the critical time point of one of the inner subformulas is returned.

%% file: sections/evaluation.tex
\section{Evaluation}
\label{sec:eval}

\newcommand{\hl}{\cellcolor{green!25}}

In this section, we evaluate the proposed \textsc{LLM-Falsifier} on standard benchmarks and conduct ablation studies to assess the effect of augmenting the prompt with additional CPS-specific information.

\subsection{Evaluation Setup}
\label{ssec:eval_setup}

\paragraph{Benchmarks}
We use benchmarks from the 2025 \textit{Applied Verification for Continuous and Hybrid Systems} competition (ARCH-COMP) falsification category, which is the most recent edition available at the time of writing~\cite{khandait2025arch}.
ARCH-COMP defines two falsification benchmark instances: \textit{Instance 1} and \textit{Instance 2}. Instance 1 allows flexible parameterization of input signals with participant-defined interpolation schemes, while Instance 2 restricts inputs to piecewise-constant signals with uniformly spaced control points and no interpolation. 
For brevity of exposition, we focus on Instance 1 for our evaluation in this paper.
The ARCH-COMP benchmarks include: \textit{Automatic Transmission} (AT)~\cite{hoxha2014benchmarks}, \textit{Neural Network Controller} (NN)~\cite{donze2010breach}, \textit{Chasing Cars} (CC)~\cite{hu2000towards}, \textit{Aircraft Ground Collision Avoidance System} (F16)~\cite{heidlauf2018verification}, and the \textit{Steam Condenser with Recurrent Neural Network Controller} (SC)~\cite{yaghoubi2019gray}. 
The only benchmark from the competition we do not include is Pacemaker (PM), as its code is not included in the public \textPsi-TaLiRo repeatability repository for ARCH-COMP.
In total, we evaluate our approach on 21 specifications across these five benchmarks, which represents a comprehensive evaluation of standard falsification tasks.
%----Added for running example:start----
The AT benchmark is the running example of Sections~\ref{sec:background} and~\ref{sec:method}: its specification $\varphi_{\mathrm{AT1}}$ appears as row AT1 in Tables~\ref{tab:archcomp-rank}--\ref{tab:model-comparison}, and the remaining AT rows are further specifications over the same model, constraining the engine speed (AT2), gear changes (AT51--AT54), and the vehicle speed under an engine-speed assumption (AT6a--AT6$_{abc}$).
%----Added for running example:end----

% 
\paragraph{Implementation Details}
We implement \textsc{LLM-Falsifier} on top of the open-source \textPsi-TaLiRo~\cite{thibeault2021psy} toolbox, which is distributed under the BSD 3-Clause License, embedding our approach as a new optimization engine. \textPsi-TaLiRo, the Python counterpart of S-TaLiRo~\cite{annpureddy2011s}, is a robustness-guided falsification framework that uses RTAMT~\cite{nivckovic2020rtamt} for quantitative robustness computation (the STL Monitor block in Figure~\ref{fig:teaser}).
We implement a recursive procedure to compute the critical time from an STL formula and trace as described in Eqs.~\eqref{eq:crit_time_start}--\eqref{eq:crit_time_end}.
Following \textPsi-TaLiRo, we use the same signal parameterization for ARCH-COMP Instance~1 benchmarks: evenly spaced control points for all input signals and Piecewise Cubic Hermite Interpolating Polynomial (\texttt{pchip}) signal interpolation, which performs shape-preserving cubic interpolation between control points.
Beyond the choice of signal parameterization, \textsc{LLM-Falsifier} has no other benchmark-specific hyperparameters.
For extracting input/output signal names to embed in the prompt (the Static Model Summarization block in Figure~\ref{fig:teaser}), we develop a parser that identifies and extracts the input signals from the system files (i.e., Simulink models with extensions \texttt{.mdl} or \texttt{.slx}), when available. Among all benchmarks, only for \textit{F16} do we specify the input names manually, as this benchmark is implemented in Python.
All experiments were run on a Linux machine with an 11th Gen Intel Core i9-11900KF CPU at 3.50\,GHz.

\paragraph{Models}
For our main falsification comparison in Section~\ref{ssec:main_comparison}, we use OpenAI \texttt{gpt-5-mini} with \texttt{high} reasoning effort as the LLM for the experiments.
For our ablation studies in Section~\ref{ssec:ablation}, in order to save on monetary costs associated with LLM API usage, we use \texttt{gpt-5-nano} with \texttt{low} reasoning effort. 
The effect of model choice is discussed in Section~\ref{ssec:model-comparison}.
OpenAI models are accessed through the OpenAI API, while the 20B open-source \texttt{gpt-oss-20b} model is run through the Hugging Face Inference Provider.
Since OpenAI does not allow users to manually adjust the \emph{temperature} in its API when using reasoning models, we use the default temperature (1.0) setting for all experiments.
For our evaluation, costs at current API rates are on the order of \$100.
\paragraph{Baselines and Evaluation Protocol}
In ARCH-COMP, due to inherent randomness in the methods, each falsification tool is evaluated on 10 independent runs per specification, with a simulation budget of up to 1500 evaluations. 
Each participant reports the \textit{Falsification Rate} (FR), defined as the number of runs out of 10 that find a counterexample, as well as the 
\emph{mean number of simulations} $\overline{S}$ when falsification succeeds.
The protocol is thus built around sample efficiency: because each simulation of a CPS model is expensive, tools are compared by how many simulations they need to find a counterexample, and the competition reports the falsification rate together with the mean and median number of simulations rather than wall-clock time. Wall-clock time is deliberately not compared, since the participants run their tools on their own machines with varying computational resources and different MATLAB/Simulink versions~\cite{khandait2025arch}. We follow the same convention and report FR and $\overline{S}$; runtime considerations for our approach are discussed in Section~\ref{sec:limitations}.
To limit API costs, we cap our simulation budget at 100 evaluations rather than 1500, which means that the falsification rate of our approach may be underestimated in the tables compared with the other tools.
We compare against a uniform random (UR) baseline and the best-performing ARCH-COMP 2025 tools from several optimization paradigms, which we describe next.

\paragraph{ARCH-COMP 2025 Baselines}
Among the eight tools reported in ARCH-COMP 2025~\cite{khandait2025arch} (including the uniform random baseline), several representative algorithmic families appear. ARIsTEO~\cite{menghi2020aristeo} uses an approximation-refinement loop, where an ARX surrogate of the CPS is learned and then iteratively refined through falsification and system identification. ATheNA~\cite{formica2024athena} is a search-based testing framework built around simulated annealing, guided by a combination of manually designed and automatically constructed fitness functions. FalCAuN~\cite{waga2020falcaun} takes a black-box checking view: it discretizes system inputs and outputs in time and value, learns a Mealy-machine abstraction of the system through active automata learning, and then uses automata-based model checking to generate counterexamples. FlexiFal~\cite{kundu2026flexifal} is a surrogate-based falsifier with two variants, NNFal and DTFal, which respectively use neural-network and decision-tree surrogates; its ARCH-COMP 2025 results were obtained with DTFal. ForeSee~\cite{zhang2021foresee} targets the scale problem that arises when a specification combines signals of different magnitudes: its QB-robustness evaluates a selected sequence of sub-formulas quantitatively and the remaining sub-formulas by Boolean satisfaction, and Monte Carlo Tree Search (MCTS) over the syntax tree of the specification chooses that sequence, with numerical optimization applied at the leaves. FReaK~\cite{bak2024hscc,bak2024atva} learns a Koopman-operator surrogate, computes reachable sets of the resulting linear model, and then uses MILP solving to identify least-robust trajectories. Finally, the \textPsi-TaLiRo competition entry uses Conjunctive Bayesian Optimization (ConBO-LS)~\cite{chotaliya2026conbo}, a Bayesian optimization method that falsifies the conjunction of all requirements of a benchmark model at once, exploiting the dependencies between requirements, and that reduces to standard Bayesian optimization when a model has a single requirement.

\subsection{Comparison with Existing Tools}
\label{ssec:main_comparison}
To rank tools, we prioritize higher falsification rate (FR) values and break ties based on lower average number of simulations $\overline{S}$. Using this ranking scheme, we compare our \textsc{LLM-Falsifier} against the best-performing ARCH-COMP 2025 baselines and a uniform random baseline (UR). 
The detailed results are in Table~\ref{tab:archcomp-rank}.
\textbf{Our \textsc{LLM-Falsifier} achieves the highest rank in 14 out of 21 specifications}.

For six of the specifications, the reasoning capabilities of the LLM allowed our tool to falsify the system in \emph{a single simulation} in every run.
For any method based on pure numerical optimization, this result would be nearly impossible as there is no information available 
at the time of the first sample.
Even the high-performance FReaK approach~\cite{bak2024hscc}, which builds a surrogate model of the CPS and reasons within it to decide on the next sample, requires an initial single random simulation to construct a surrogate model.

\begin{table}[t]
\centering
\caption{Falsification results on the ARCH-COMP falsification benchmarks (Instance~1). For each specification, \emph{ARCH-COMP Best} is the best-performing ARCH-COMP 2025 tool, \emph{Rank} is the position of \textsc{LLM-Falsifier} when inserted into the ARCH-COMP 2025 ranking, and \emph{UR} is uniform random sampling. Green cells mark the best result per row, an asterisk (*) indicates a tie, and a dash indicates that no run found a counterexample.}
\label{tab:archcomp-rank}
\setlength{\tabcolsep}{3.5pt} % Adjust this value to increase/decrease spacing
\begin{tabular}{l|crr|crr|rr}
\toprule
~ & \multicolumn{3}{c|}{ARCH-COMP Best} & \multicolumn{3}{c|}{Ours} & \multicolumn{2}{c}{UR} \\
Spec & Tool & FR & $\overline{S}$ & Rank & FR & $\overline{S}$ & FR & $\overline{S}$ \\
\midrule
AT1 & FReaK & 10 & 4.8 & \hl 1st & 10 & 1.0 & 0 & - \\
AT2 & FReaK & 10 & 2.1 & \hl 1st & 10 & 1.0 & 10 & 7.6 \\
AT51 & \hl FReaK & 10 & 8.7 & - & 0 & - & 1 & 923.0 \\
AT52 & \hl FReaK & 10 & 1.3 & \hl 1st* & 10 & 1.3 & 10 & 4.1 \\
AT53 & \hl FReaK & 10 & 1.1 & 2nd & 10 & 2.3 & 10 & 18.6 \\
AT54 & \hl FReaK & 10 & 2.4 & - & 0 & - & 3 & 932.0 \\
AT6a & FReaK & 10 & 7.4 & \hl 1st & 10 & 3.3 & 10 & 74.4 \\
AT6b & FReaK & 10 & 6.2 & \hl 1st & 10 & 4.7 & 10 & 251.3 \\
AT6c & FReaK & 10 & 5.9 & \hl 1st & 10 & 4.8 & 10 & 185.2 \\
AT6$_{abc}$ & FReaK & 10 & 6.4 & \hl 1st & 10 & 3.1 & 10 & 58.8 \\
\midrule
NN & \hl FReaK & 10 & 2.0 & 6th & 1 & 97.0 & 10 & 38.6 \\
NN$\beta$ & \hl FReaK & 10 & 30.9 & - & 0 & - & 0 & - \\
NNx & FReaK & 10 & 192.3 & \hl 1st & 10 & 1.0 & 0 & - \\
\midrule
CC1 & FReaK & 10 & 3.6 & \hl 1st & 10 & 1.0 & 10 & 10.4 \\
CC2 & FReaK & 10 & 3.0 & \hl 1st & 10 & 1.0 & 10 & 15.4 \\
CC3 & FReaK & 10 & 5.7 & \hl 1st & 10 & 1.7 & 10 & 77.9 \\
CC4 & \hl FReaK & 10 & 176.7 & 3rd & 5 & 58.6 & 0 & - \\
CC5 & ARIsTEO & 10 & 31.4 & \hl 1st & 10 & 17.1 & 10 & 28.5 \\
CCx & FReaK & 10 & 110.5 & \hl 1st & 10 & 9.0 & 7 & 338.1 \\
\midrule
F16 & \hl FReaK & 10 & 1.0 & \hl 1st* & 10 & 1.0 & 0 & - \\
\midrule
SC & \hl FReaK & 10 & 45.1 & - & 0 & - & 0 & - \\
\bottomrule
\end{tabular}
\end{table}

\subsection{Ablation Studies}
\label{ssec:ablation}

To justify the choice of our prompt design decisions, we performed an ablation study comparing the four progressive meta-prompt variants defined in Section~\ref{sec:signal_names}, namely MP1 through MP4.

Table~\ref{tab:main-ablation-study} summarizes the ablation study results. The same ranking criterion described before in Section~\ref{ssec:main_comparison} was used to assess performance. 
Across most benchmarks, adding more context tends to improve performance, with MP4 achieving the best results.
Providing semantically meaningful model-specific context, particularly critical-time information, helps the LLM focus its search and identify counterexamples more efficiently. 

The largest performance gains were observed for the Automatic Transmission (AT) benchmarks. 
We believe this improvement stems from the fact that the input signal names closely correspond to their physical meaning. Specifically, in the AT case, the inputs are \texttt{throttle} and \texttt{brake}, while the outputs include \texttt{speed}, \texttt{rpm}, and \texttt{gear}, which are directly referenced in the specifications. This clear semantic alignment allows the LLM to reason more effectively about cause-and-effect relationships, making falsification easier. 
The running example illustrates this: with the baseline prompt MP1, which lists the ten input dimensions only by index and reports only the scalar robustness, \texttt{gpt-5-nano} needed $25.2$ simulations on average to falsify AT1 and failed in one of ten runs.
%----Added for running example:start----
Once the prompt names the inputs \texttt{throttle} and \texttt{brake} and the output \texttt{speed} and states the specification $\mathbf{G}_{[0,20]}(\texttt{speed}\le 120)$ (MP2), the same model succeeds in every run after $1.4$ simulations on average, since it can infer that high throttle and no braking maximize the speed instead of discovering this relation by trial and error (see the reasoning trace in Section~\ref{ssec:explainability}).
%----Added for running example:end----

The failure cases in AT correspond to specifications AT51 and AT54. Together with the related specifications AT52 and AT53, they are defined over the \texttt{gear} signal, which takes discrete values. This discreteness causes the robustness measure to plateau at fixed levels (e.g., 0.5), making these specifications inherently more difficult to falsify compared to the other AT cases.
A possible future improvement could extract more graybox model information in the Static Model Summarization block from Figure~\ref{fig:teaser}, for example, explaining how RPM determines gear to improve the falsification search.

%% Ablation study table
\begin{table}[t]
\centering
\caption{Ablation study with \texttt{gpt-5-nano} (low reasoning) measuring the impact of each contextual element in the meta-prompt (MP1--MP4, Section~\ref{sec:signal_names}). Green cells mark the best result per row.}
\label{tab:main-ablation-study}
\setlength{\tabcolsep}{3.5pt} % Adjust this value to increase/decrease spacing
\begin{tabular}{l|rr|rr|rr|rr}
\toprule
~ & \multicolumn{2}{c|}{MP1} & \multicolumn{2}{c|}{MP2} & \multicolumn{2}{c|}{MP3} & \multicolumn{2}{c}{MP4} \\
Spec & FR & $\overline{S}$ & FR & $\overline{S}$ & FR & $\overline{S}$ & FR & $\overline{S}$ \\
\midrule
AT1 & 9 & 25.2 & 10 & 1.4 & 10 & 1.4 & \hl 10 & \hl 1.2 \\
AT2 & 10 & 1.5 & \hl 10 & \hl 1.0 & \hl 10 & \hl 1.0 & \hl 10 & \hl 1.0 \\
AT51 & 0 & - & 0 & - & 0 & - & 0 & - \\
AT52 & 10 & 7.3 & 10 & 8.7 & \hl 10 & \hl 4.6 & 10 & 6.5 \\
AT53 & 10 & 7.1 & \hl 10 & \hl 6.7 & 10 & 7.7 & 10 & 6.8 \\
AT54 & 0 & - & 0 & - & 0 & - & 0 & - \\
AT6a & 7 & 42.6 & 9 & 33.9 & 10 & 15.3 & \hl 10 & \hl 9.7 \\
AT6b & 5 & 60.2 & 9 & 50.7 & 10 & 14.2 & \hl 10 & \hl 9.3 \\
AT6c & 7 & 61.1 & 10 & 13.3 & 10 & 7.8 & \hl 10 & \hl 6.6 \\
AT6$_{abc}$ & 9 & 43.8 & 10 & 18.0 & 10 & 12.3 & \hl 10 & \hl 11.3 \\
\midrule
NN & 0 & - & 0 & - & 0 & - & 0 & - \\
NN$\beta$ & 0 & - & 0 & - & 0 & - & 0 & - \\
NNx & \hl 10 & \hl 1.0 & \hl 10 & \hl 1.0 & \hl 10 & \hl 1.0 & \hl 10 & \hl 1.0 \\
\midrule
CC1 & 10 & 11.4 & \hl 10 & \hl 1.0 & 10 & 1.1 & \hl 10 & \hl 1.0 \\
CC2 & 10 & 2.6 & \hl 10 & \hl 1.0 & 10 & 1.1 & \hl 10 & \hl 1.0 \\
CC3 & 10 & 14.7 & \hl 10 & \hl 1.0 & \hl 10 & \hl 1.0 & \hl 10 & \hl 1.0 \\
CC4 & 0 & - & \hl 3 & \hl 32.0 & 3 & 50.0 & 0 & - \\
CC5 & \hl 10 & \hl 12.0 & 5 & 59.0 & 4 & 49.5 & 7 & 29.0 \\
CCx & 1 & 43.0 & 1 & 49.0 & 1 & 76.0 & \hl 1 & \hl 20.0 \\
\midrule
F16 & 10 & 3.3 & 10 & 4.1 & \hl 10 & \hl 1.9 & 10 & 3.2 \\
\midrule
SC & 0 & - & 0 & - & 0 & - & 0 & - \\
\bottomrule
\end{tabular}
\end{table}

\subsection{Effect of Model Choice}
\label{ssec:model-comparison}

\begin{figure*}[!t]
  \centering
  \includegraphics[width=0.99\linewidth]{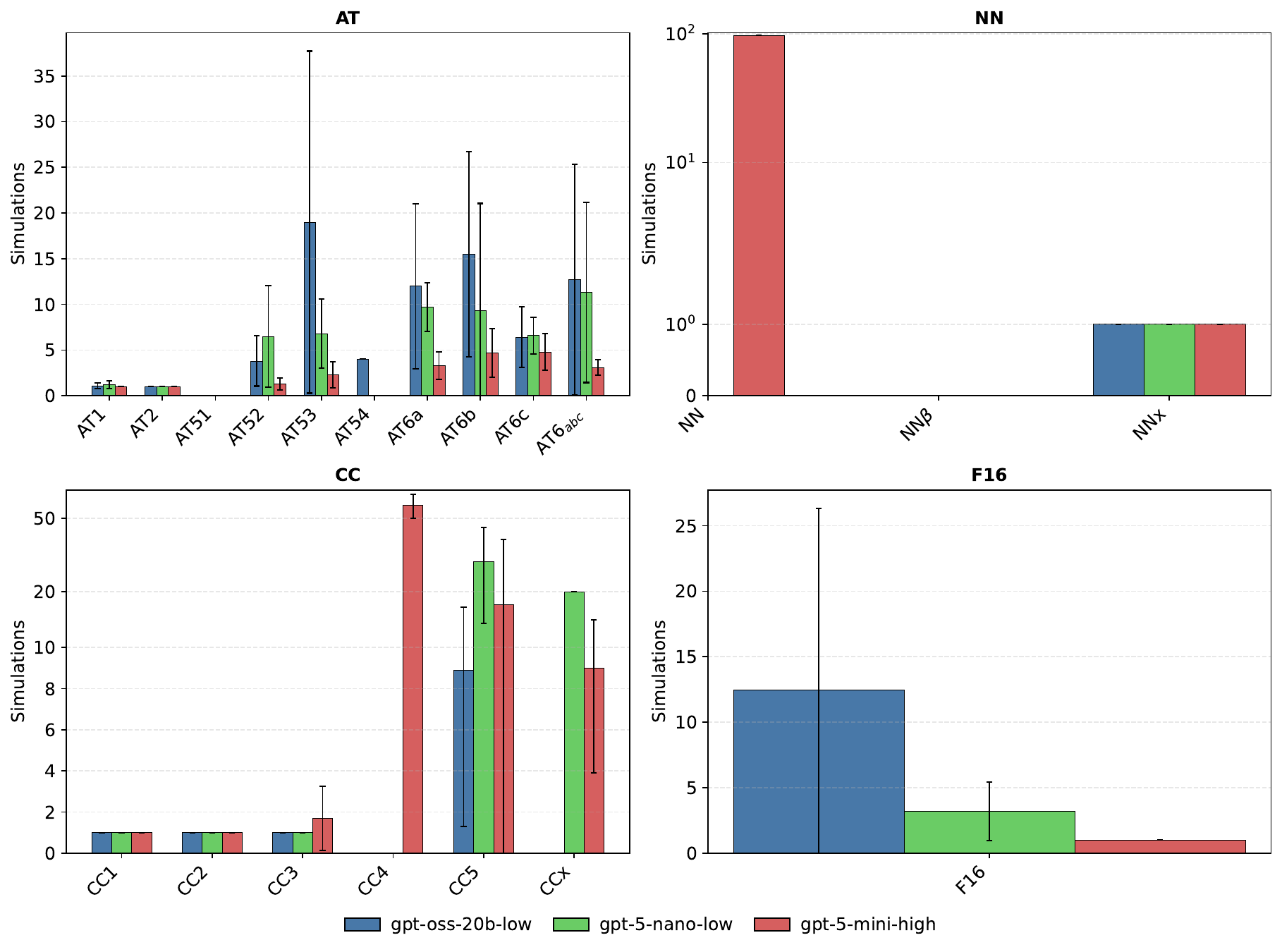}
  \caption{Mean number of simulations required to find a falsifying input for the three LLMs on all specifications except SC, which no model falsified. Bars show the mean $\pm$ standard deviation over successful falsification runs; the NN panel uses a logarithmic scale. The more capable the model, the lower the mean and standard deviation on almost all specifications. Missing bars indicate that the model found no counterexample within the budget.}
  \Description{Grouped bar chart over all benchmark specifications except SC, comparing the mean number of simulations for gpt-oss-20b, GPT-5-nano, and GPT-5-mini. GPT-5-mini generally achieves the lowest means and is the only model with bars on the hardest specifications such as NN, CC4, and CCx.}
  \label{fig:models-ablation}
\end{figure*}

In addition to meta-prompt configurations, we study the impact of the underlying model choice across all evaluated specifications. Table~\ref{tab:model-comparison} provides the complete falsification rate (FR) and mean number of simulations ($\overline{S}$) for three models: \texttt{gpt-oss-20b}, \texttt{gpt-5-nano}, and \texttt{gpt-5-mini}. To visually compare their search efficiency, Figure~\ref{fig:models-ablation} plots the mean simulations and standard deviation across all successful falsification runs for each specification.

\begin{table}[t]
\centering
\caption{Effect of model choice with the full prompt MP4. Larger models with more reasoning effort perform better, although the open-source option is competitive. Green cells mark the best result per row.}
\label{tab:model-comparison}
\setlength{\tabcolsep}{3.5pt}
\begin{tabular}{l|rr|rr|rr}
\toprule
~ & \multicolumn{2}{c|}{\texttt{gpt-oss-20b}} & \multicolumn{2}{c|}{\texttt{gpt-5-nano}} & \multicolumn{2}{c}{\texttt{gpt-5-mini}} \\
~ & \multicolumn{2}{c|}{(low reas.)} & \multicolumn{2}{c|}{(low reas.)} & \multicolumn{2}{c}{(high reas.)} \\
Spec    & FR & $\overline{S}$ & FR & $\overline{S}$ & FR & $\overline{S}$ \\
\midrule
AT1     & 10 & 1.1 & 10 & 1.2 & \hl 10 & \hl 1.0 \\
AT2     & \hl 10 & \hl 1.0 & \hl 10 & \hl 1.0 & \hl 10 & \hl 1.0 \\
AT51    & 0 & - & 0 & - & 0 & - \\
AT52    & 10 & 3.8 & 10 & 6.5 & \hl 10 & \hl 1.3 \\
AT53    & 9 & 19.0 & 10 & 6.8 & \hl 10 & \hl 2.3 \\
AT54    & \hl 1 & \hl 4.0 & 0 & - & 0 & - \\
AT6a    & 10 & 12.0 & 10 & 9.7 & \hl 10 & \hl 3.3 \\
AT6b    & 10 & 15.5 & 10 & 9.3 & \hl 10 & \hl 4.7 \\
AT6c    & 10 & 6.4 & 10 & 6.6 & \hl 10 & \hl 4.8 \\
AT6$_{abc}$  & 10 & 12.7 & 10 & 11.3 & \hl 10 & \hl 3.1 \\
\midrule
NN      & 0 & - & 0 & - & \hl 1 & \hl 97.0 \\
NN$\beta$ & 0 & - & 0 & - & 0 & - \\
NNx     & \hl 10 & \hl 1.0 & \hl 10 & \hl 1.0 & \hl 10 & \hl 1.0 \\
\midrule
CC1     & \hl 10 & \hl 1.0 & \hl 10 & \hl 1.0 & \hl 10 & \hl 1.0 \\
CC2     & \hl 10 & \hl 1.0 & \hl 10 & \hl 1.0 & \hl 10 & \hl 1.0 \\
CC3     & \hl 10 & \hl 1.0 & \hl 10 & \hl 1.0 & 10 & 1.7 \\
CC4     & 0 & - & 0 & - & \hl 5 & \hl 58.6 \\
CC5     & \hl 10 & \hl 8.9 & 7 & 29.0 & 10 & 17.1 \\
CCx     & 0 & - & 1 & 20.0 & \hl 10 & \hl 9.0 \\
\midrule
F16     & 9 & 12.4 & 10 & 3.2 & \hl 10 & \hl 1.0 \\
\midrule
SC      & 0 & - & 0 & - & 0 & - \\
\bottomrule
\end{tabular}
\end{table}

Across most benchmarks, larger models with higher reasoning effort consistently yield higher falsification rates and require fewer simulations. On the Automatic Transmission (AT) benchmarks, \texttt{gpt-5-mini} with high reasoning consistently requires the lowest mean number of simulations to discover counterexamples, while the open-source \texttt{gpt-oss-20b} remains highly competitive, often achieving falsification in a comparable number of simulations to \texttt{gpt-5-nano}. On simpler or highly semantic tracking benchmarks like CC3 and CC5, \texttt{gpt-oss-20b} is very efficient, sometimes finding counterexamples in fewer average simulations than \texttt{gpt-5-mini}. The advantage of larger models is particularly evident in difficult specifications:
\begin{itemize}
    \item \textbf{Neural Network Controller (NN)}: Only \texttt{gpt-5-mini} (high reasoning) was able to falsify the NN specification, and only in 1 of 10 runs within the 100-simulation budget.
    \item \textbf{Chasing Cars (CC4, CCx)}: On CC4, only \texttt{gpt-5-mini} achieved a non-zero falsification rate (FR = 5). Similarly, on CCx, \texttt{gpt-5-mini} succeeded in all 10 runs with a mean of only 9.0 simulations, whereas \texttt{gpt-5-nano} succeeded in a single run and \texttt{gpt-oss-20b} found no counterexample.
\end{itemize}
These results suggest that complex temporal specifications require advanced logical and negation reasoning, which is stronger in larger models with dedicated reasoning steps.
At the same time, the open-source \texttt{gpt-oss-20b} model exhibits competitive performance on several benchmarks. This indicates that \textsc{LLM-Falsifier} generalizes across different models and stands to benefit further as language models continue to improve.

\subsection{Explainability}
\label{ssec:explainability}
To understand how the LLM constructs samples that frequently falsify the specifications, we inspected its explicit reasoning output (using the open-source \texttt{gpt-oss-20b}~\cite{agarwal2025gpt} model, since full reasoning traces of the GPT-5 models are not exposed through the API). We return to the running example: the prompt in Figure~\ref{fig:meta-prompt} with specification $\varphi_{\mathrm{AT1}} = \mathbf{G}_{[0,20]}(\texttt{speed} \leq 120)$. In this case the model immediately produced a falsifying sample on the first attempt, that is, from the meta-prompt alone, with an empty history section. We observed the following reasoning output:
\begin{quote}
\texttt{"Need high throttle and low brake to exceed 120 speed. Use high throttle 100, low brake 0."}
\end{quote}
The output sample had $100$ throttle and $0$ brake at all time points, i.e., the point $(100,\dots,100,\,0,0,0)$ in the ten-dimensional input space of Section~\ref{sec:background}.
This closes the loop of Figure~\ref{fig:teaser} for the running example: the Static Model Summarization step supplied the input names \texttt{throttle} and \texttt{brake} from the Simulink model and the output name \texttt{speed} from the specification; the LLM negated the specification (the speed must exceed $120$ at some time in $[0,20]$) and used the physical meaning of the names to conclude that full throttle and no braking maximize the speed; the simulator produced a trajectory whose speed exceeds $120$ within the window (the counterexample trace of Figure~\ref{fig:at1-example}); and the STL Monitor returned a negative robustness, so the search terminated after a single simulation.
This is the behavior behind the AT1 rows of Tables~\ref{tab:archcomp-rank}--\ref{tab:model-comparison}, and it is unattainable for a purely numerical optimizer, which has no information before its first sample.
Based on our observations, the LLM's falsification reasoning generally follows three stages: (1) negate the STL formula to identify a violation strategy, (2) propose control actions that drive the system toward violation, and (3) refine action magnitudes based on previous examples or feedback.

This behavior is not limited to simple predicates and extends to more complex temporal logic structures. 
For example, consider specification AT6a: $(\mathbf{G}_{[0,30]}(\texttt{rpm} \leq 3000)) \rightarrow (\mathbf{G}_{[0,4]}(\texttt{speed} \leq 35))$. 
For this STL specification, the model generated the following text as part of its reasoning process:
\begin{quote}
\texttt{"To violate, need antecedent true but consequent false. So need rpm always <=3000 for t 0-30, but speed >35 at some t <=4. So we need high speed early while keeping rpm low. From samples, ..."}
\end{quote}
Such reasoning chains demonstrate effective logical reasoning capabilities of the LLM. 
The model correctly negates the STL implication operator and formulates a valid falsification strategy. It also shows that the LLM can interpret STL specifications directly in symbolic form, without requiring translation into natural language.

%% file: sections/limitations.tex
\section{Limitations}
\label{sec:limitations}

Our approach has several limitations. First, it inherits the computational cost and latency of modern reasoning-enabled LLMs. Following the ARCH-COMP reporting convention, our evaluation tables report \textit{Falsification Rate} (FR) and the mean number of simulations $\overline{S}$ rather than wall-clock execution time. This convention exists because wall-clock time is not comparable across the participating tools: their results are collected on different machines, and the tools span very different computational paradigms, from surrogate-model training that benefits from GPU acceleration to CPU-bound MILP solving, automata learning, and simulated annealing. Sample efficiency therefore does not necessarily translate into lower runtime for any tool, including ours. Wall-clock time depends on several external factors, including the machine or server used to run the LLM, the machine used to execute the remaining falsification pipeline and simulations, and the particular benchmark and specification. To provide a sense of runtime, the mean time per iteration for the AT benchmark in our experiments was 4.4~s for \texttt{gpt-oss-20b} (low reasoning), 10.9~s for \texttt{gpt-5-nano} (low reasoning), and 81.8~s for \texttt{gpt-5-mini} (high reasoning). Most specifications falsified by our approach required fewer than 10 iterations. Nevertheless, API usage was materially more expensive than classical falsification heuristics, and LLM inference can add noticeable delay to each iteration. Although LLM-based search can substantially reduce the number of simulator calls, this reduction does not automatically imply lower wall-clock time or lower monetary cost. As a result, our method is currently most attractive in settings where simulator evaluations are expensive and sample efficiency matters more than raw inference cost.

Second, the method appears to benefit most when the prompt exposes semantically meaningful structure that an LLM can exploit. In benchmarks such as Automatic Transmission, input and output names like \texttt{throttle}, \texttt{brake}, \texttt{speed}, and \texttt{rpm} provide strong causal cues. In contrast, performance is weaker on benchmarks where this semantic connection is indirect, missing, or less informative, and on cases with discrete or plateaued robustness landscapes such as some gear-based specifications. This suggests that the effectiveness of LLM-guided falsification may depend on how naturally the CPS and specification can be rendered into a semantically informative prompt.

Finally, our explainability evidence is preliminary. Because we did not have access to full reasoning traces for the proprietary model used in the main experiments, the qualitative analysis of intermediate reasoning relied on a different open-source model. These examples are useful for illustrating plausible reasoning patterns, but they should not be interpreted as direct evidence of the internal reasoning process of the main model used for the strongest quantitative results.

%% file: sections/conclusion.tex
\section{Conclusion and Discussion}
% \section{Conclusion}
\label{sec:conclusion}

This paper introduced and evaluated \textsc{LLM-Falsifier}, which is, to our knowledge, the first robustness-guided falsification framework that uses a large language model as the optimizer.
Our results show that LLMs can directly optimize STL robustness values and become substantially more effective when the search loop exposes semantically meaningful context, including natural-language variable names, output trajectories, and critical-time witnesses.
On the ARCH-COMP falsification benchmarks, \textsc{LLM-Falsifier} outperforms established falsification tools based on diverse optimization strategies, from surrogate-based and Bayesian optimization to search-based testing, on 14 of 21 specifications.
These results suggest that LLMs are not only generic prompt optimizers, but can also serve as competitive search procedures for formal reasoning tasks when the optimization loop is expressed in a semantically rich form.
On six specifications, the LLM found a falsifying sample on the \emph{first} try in every run, highlighting the potential value of language-mediated reasoning in sample-efficient search.

Several directions could extend this work. The prompt-based formulation naturally leaves room for additional graybox model information (e.g., dynamic values of internal signals) in the Static Model Summarization step, which could further improve search efficiency. Another promising avenue is a hybrid approach that combines LLM-based exploration with classical numeric optimization for fine-grained exploitation.

By combining formal robustness targets with structured semantic prompts, our framework suggests a general recipe for optimization problems requiring both scalar feedback and semantic context.